\documentclass[10pt]{iopart}
\usepackage[utf8]{inputenc}
\usepackage{graphicx}
\usepackage{booktabs}
\usepackage{multirow}
\usepackage{cite}
\usepackage{amssymb}
\usepackage[hidelinks]{hyperref}

\usepackage[textsize=footnotesize]{todonotes}

\begin{document}
\frenchspacing

\title{Impact of dataset size and long-term ECoG-based BCI usage on deep learning decoders performance}

% \author{Anonymous authors}
\author{Maciej Śliwowski$^{1, 2}$, Matthieu Martin$^1$, Antoine Souloumiac$^2$, Pierre Blanchart$^2$ and Tetiana Aksenova$^1$}

\address{1 Univ. Grenoble Alpes, CEA, LETI, Clinatec, F-38000 Grenoble, France}
\address{2 Université Paris-Saclay, CEA, List, F-91120, Palaiseau, France}
\ead{\mailto{tetiana.aksenova@cea.fr}}
\vspace{10pt}
\begin{indented}
\item[] August 2022
\end{indented}

\begin{abstract}
\textit{Objective.}
In brain-computer interfaces (BCI) research, recording data is time-consuming and expensive, which limits access to big datasets. This may influence the BCI system performance as machine learning methods depend strongly on the training dataset size. Important questions arise: taking into account neuronal signal characteristics (e.g., non-stationarity), can we achieve higher decoding performance with more data to train decoders? What is the perspective for further improvement with time in the case of long-term BCI studies? In this study, we investigated the impact of long-term recordings on motor imagery decoding from two main perspectives: model requirements regarding dataset size and potential for patient adaptation.

\textit{Approach.}
We evaluated the multilinear model and two deep learning (DL) models on a long-term  BCI \& Tetraplegia NCT02550522 clinical trial dataset containing 43 sessions of ECoG recordings performed with a tetraplegic patient. In the experiment, a participant executed 3D virtual hand translation using motor imagery patterns. We designed multiple computational experiments in which training datasets were increased or translated to investigate the relationship between models' performance and different factors influencing recordings.

\textit{Main results.}
For all tested decoders, our analysis showed that adding more data to the training dataset may not instantly increase performance for datasets already containing 40 minutes of the signal. DL decoders showed similar requirements regarding the dataset size compared to the multilinear model while demonstrating higher decoding performance. Moreover, high decoding performance was obtained with relatively small datasets recorded later in the experiment, suggesting motor imagery patterns improvement and patient adaptation during the long-term experiment. Finally, we proposed UMAP embeddings and local intrinsic dimensionality as a way to visualize the data and potentially evaluate data quality.

\textit{Significance.}
DL-based decoding is a prospective approach in BCI which may be efficiently applied with real-life dataset size. Patient–decoder co-adaptation is an important factor to consider in long-term clinical BCI.
% Machine learning models evaluated in the study could not achieve higher performance when using more data. Instead, the highest cosine similarity was achieved with relatively small datasets recorded later in the clinical trial emphasizing the importance of patient adaptation and motor imagery patterns improvement.
\end{abstract}
\vspace{2pc}

\ioptwocol
\section{Introduction}

Permanent motor deficits as a result of a spinal cord injury (SCI) affect hundreds of thousands of people worldwide each year (12,000 people each year just in the United States \cite{hachem_assessment_2017}). In this case, the motor cortex is preserved, but neuronal signals can no longer be transmitted to the muscles. Then, the use of a brain-computer interface (BCI), which enables interacting with an effector by thought, could enable these patients to regain a certain autonomy in everyday life. For example, motor imagery based BCI has been used for the control of prostheses or exoskeletons of upper limbs \cite{edelman2019noninvasive, hochberg2012reach, wodlinger2014ten}, lower limbs \cite{lopez2016control, he2018brain, kwak2015lower, collinger2013high} and four limbs \cite{benabid2019exoskeleton} in subjects with paraplegia or tetraplegia following an SCI. In this study, we focus on electrocorticography (ECoG)-based motor BCIs, promising tools that may enable continuous 3D hand trajectory decoding for neuroprosthesis control while reducing the risk of implantation compared to more invasive approaches \cite{volkova_ecog_review}.

BCIs record neuronal activity and decode it into control commands for effectors. Decoders are generally trained using machine learning algorithms in a supervised manner. In the vast majority of studies, the training dataset is strongly restricted due to limited access to recordings. At the same time, dataset size is an important factor in machine learning analysis and can influence overall system performance drastically. In contrast to recent computer vision and natural language processing studies \cite{kaplan, Hoiem2021LearningCF, rosenfeld2020a}, the optimal quantity of training data, i.e., the quantity at which decoder's performance reaches a plateau for a given application, is rarely studied for BCI \cite{perdikis2020brain}. Especially, learning curves, providing insight into the relationship between model performance and training set size, are rarely presented. Learning curves can be used for model selection, decreasing the computational load of model training, or estimating the theoretical influence of adding more data to training datasets \cite{viering_learning_curves_review}. The last point is particularly important in BCI, considering limited access to datasets recorded with humans. Without knowing the relationship between system performance and dataset size, it is hard to determine the strategy to improve the accuracy of decoders: increase the amount of training data or increase the capacity of the models. In the case of ECoG-based motor BCI, most models have a limited capacity. The decoders used are Kalman filters \cite{silversmith_plug-and-play_2020, pistohl_decoding_2012} and mostly variants of linear models \cite{liang_decoding_2012, nakanishi_prediction_2013, nakanishi_mapping_2017, flamary_decoding_2012, weixuan_chen_logistic_weighted_2014, bundy_decoding_2016, eliseyev_recursive_2017}. In most of these studies, decoder optimization has been carried out on databases containing a few minutes or tens of minutes of the signal. This results in usable models but does not provide any information on the performance gain that could be achieved with more data, nor does it compare the data quantity/performance relationship between several decoders.

Model characteristics and learning curves are not the only factors influencing decoders' performance in the case of BCI. The human ability to generate distinct brain signal patterns is crucial for a BCI system to work. Research in recent years has focused mainly on the development of increasingly efficient decoders, for example DL \cite{lawhern2018eegnet, elango2017sequence, schirrmeister2017deep, zhang2019novel, bashivan2015learning, pan2018rapid, rashid2020electrocorticography, xie2018decoding, du_decoding_2018, elango_sequence_2017, pan_rapid_2018, rashid_electrocorticography_2020, xie_decoding_2018, sliwowski2022decoding} rather than on patient learning or co-adaptation \cite{millan2004need, wolpaw2002brain}, even though several studies have shown the crucial importance of patient learning \cite{carmena2013advances, chavarriaga2017heading, lotte2013flaws, orsborn2014closed, lotte_long_term}. Thanks to recording device developments and clinical trial advances, long-term studies of chronic BCI enable recording bigger datasets than ever before. Current techniques for recording brain activity, such as the ElectroCorticoGram (ECoG), provide stable recordings for at least 2 years \cite{nurse2017consistency}. It offers the possibility to train and test a decoder over several months. It also enables studying potential patient learning and provides insight into the optimal quantity of data necessary to get the best out of a decoder. These questions have largely been put aside \cite{perdikis2020brain}.

Closed-loop learning allows for short-term patient-model co-adaptation through the visual feedback received by the patient. This feedback leads to a modification of the brain activity and has shown capabilities for improving the control of neuroprostheses \cite{cunningham2011closed, jarosiewicz2013advantages, sitaram2017closed, orsborn2014closed, shanechi2017rapid}. Nevertheless, motor learning is a process that takes place in the short term and in the long term \cite{dayan2011neuroplasticity, krakauer2019motor}. This long-term learning is little studied in BCI and most studies in humans are limited to a few sessions ($< 15$) \cite{holz2013brain, hohne2014motor, meng2016noninvasive, leeb2015towards} to show that a fast and efficient calibration of the proposed decoders is possible. Several studies with a larger number of sessions ($> 20$) were nevertheless carried out: \cite{neuper2003clinical, mcfarland2010electroencephalographic, collinger2013high, hochberg2006neuronal, ajiboye2017restoration, wodlinger2014ten, perdikis2018cybathlon, wolpaw2004control, moly2022adaptive, lotte_long_term}. Some have focused on patient learning \cite{neuper2003clinical, mcfarland2010electroencephalographic, collinger2013high, leeb2015towards, perdikis2018cybathlon, lotte_long_term} by seeking an improvement in performance coming from changes in the signal or the characteristics extracted from it. The last point is required to distinguish between performance improvement due to patient learning, increased data available for decoder optimization, or changes in the experimental environment \cite{perdikis2020brain, lotte2018defining}.

This study investigated the relationship between BCI decoders' performance predicting 3D upper-limb movements from ECoG signals and the training dataset size used to optimize model parameters. Learning curves obtained in different offline computational experiments showed that multilinear and DL models saturate at a similar amount of data, between 30 and 90 minutes of ECoG signal, depending on the scenario and hand. Moreover, learning curves revealed characteristics that were unlikely caused by just the dataset increase. Extended analysis using unsupervised ML methods showed dataset characteristic changes with time, suggesting that long-term patient learning may play an important role in achieving higher BCI performance. This kind of analysis was possible thanks to the access to a rare database of ECoG signals \cite{moly2022adaptive} containing imagined hand movements performed by a tetraplegic patient to control upper-limb 3D translation in a virtual environment. The dataset contains 43 sessions recorded over 9 months (approximately 6 hours of data for each hand).

\section{Methods}

\subsection{Clinical trial and patient} \label{sec:dataset}
The data was recorded and analyzed as a part of the "BCI and Tetraplegia" (ClinicalTrials.gov identifier: NCT02550522) clinical trial, which was approved by the Agency for the Safety of Medicines and Health Products (Agence nationale de sécurité du médicament et des produits de santé---ANSM) with the registration number: 2015-A00650-49 and the ethical Committee for the Protection of Individuals (Comité de Protection des Personnes---CPP) with the registration number: 15-CHUG-19.

The participant was a 28-year-old right-handed man following tetraplegia after a C4-C5 spinal cord injury. He had residual control over upper limbs with American Spinal Injury Association Impairment (ASIA) scores of 4 (right hand), 5 (left hand) at the level of the elbow, and 0 (right hand), 3 (left hand) at the extensors of the wrist. All motor functions below were completely lost (ASIA score of 0). \cite{benabid_exoskeleton_2019}

Two WIMAGINE implants \cite{mestais_wimagine_2015}, recording ECoG signal at 586 Hz sampling rate, were implanted above left and right primary motor and sensory cortex responsible for upper limb movements. The implants consisted of an $8 \times 8$ electrode's grid. Due to the data transfer limit, only 32 electrodes (organized on a chessboard-like grid) were used.

The data recordings used in this study started after 463 days post-implantation. The subject was already experienced in the BCI setup as the clinical trial experiments began shortly after the surgery. During the clinical trial, the participant gradually learned how to control the BCI, starting by using discrete/1D effectors and finally achieving control of up to 8D movements in one experimental session.
% \begin{figure}
% 	\centering \includegraphics[width=0.6\columnwidth]{images/fig02.png}
% 	\caption{Position of the implants (green) and electrodes (red) on an MRI reconstruction of the patient's brain.} \label{fig:implants-brain}
% \end{figure}

\subsection{Data and experimental paradigm}
\begin{figure}
    \centering
    \includegraphics[width=\columnwidth]{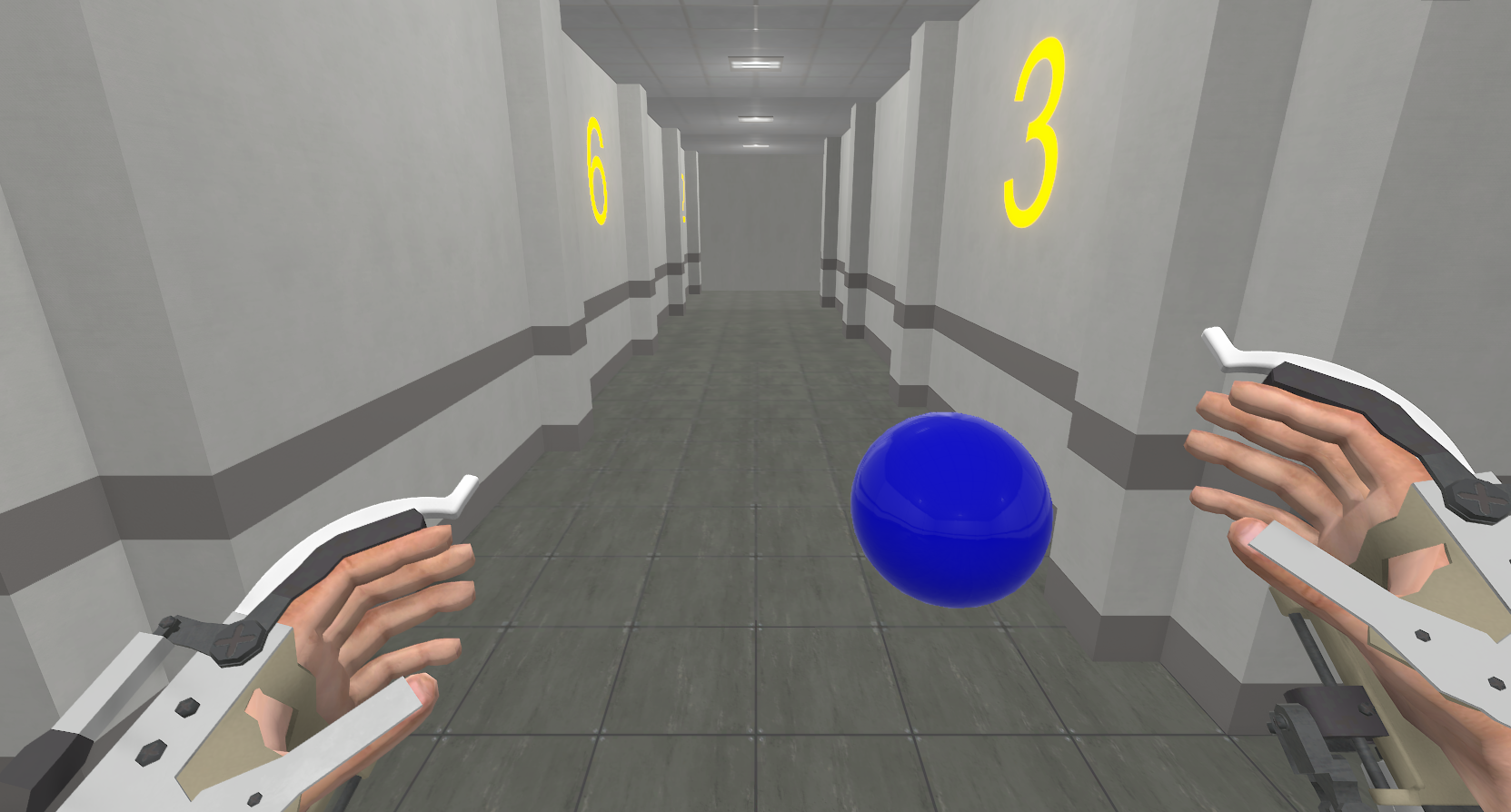}
    \caption{Screenshot from the virtual environment. The patient was asked to reach the blue sphere with his right hand.}
    \label{fig:virtual-env}
\end{figure}

The dataset analyzed in this study contains 43 experimental sessions in which tetraplegic patient was asked to perform motor imagery tasks in order to move virtual exoskeleton effectors (see the virtual environment in figure \ref{fig:virtual-env}). In particular, the patient used an MI strategy in which he repeatedly imagined/attempted fingers, hands, and arm movements to control 8 dimensions (3D left and right hand translation and 1D left and right wrist rotation). In every trial, the patient's goal was to reach the target displayed on the screen, one after another, without returning to the center position. \cite{moly2022adaptive}

During the experimental sessions, 1 out of 5 states (idle state, left hand translation, right hand translation, left wrist rotation, right wrist rotation) was decoded from the recorded ECoG signal with a multilinear gate model. Accordingly to the gate predictions, an appropriate multilinear expert was selected to provide a trajectory of hand movement or direction of wrist rotation. For further analysis, we selected only left and right hand translation datasets.

Multilinear model parameters were optimized online during the recordings using recursive exponentially weighted n-way partial least squares (REW-NPLS) \cite{eliseyev2017recursive}. Models were trained on the first six sessions, further referred to as the calibration dataset. For the next 37 sessions, models' weights were fixed and used for the performance evaluation. In our computational experiments, we concatenated calibration and test sessions to perform offline simulations in different scenarios, studying the dataset and model characteristics in-depth. Datasets sizes are reported in table \ref{tab:data}.

\begin{table}
	\caption{Datasets size in the number of trials and length of the recordings.}
	\label{tab:data}
	\centering
    \begin{tabular}{lrr}
    \toprule
    {} &  Left hand &  Right hand \\
    \midrule
    Trials         &        811 &         756 \\
    Duration [min] &        300 &         284 \\
    \bottomrule
    \end{tabular}
\end{table}

% \begin{figure}
% 	\centering \includegraphics[width=\columnwidth]{images/fig01.png}
% 	\caption{A) Screenshot from the virtual environment displaying the hand of the avatar and the target. B) Visualization of the axes of the coordinate system of the virtual avatar.} \label{fig:experiment-ve}
% \end{figure}

% Targets were placed in 28 (LH) and 23 (RH) positions during the experiments (see \ref{app:targets-positions} for targets positions visualization). The number of trials and minutes of the recorded signal are given for both hands and the calibration and test datasets in Table \ref{tab:data}.

\subsection{Preprocessing and feature extraction}

Raw ECoG signal was processed with a feature extraction pipeline creating time-frequency representation. Continuous complex wavelet transform was used with 15 Morlet wavelets with central frequencies in the range of 10-150 Hz (10 Hz interval). Every 100 ms, one second of signal (90\% overlap) was selected and convolved with the set of wavelets coefficients. Then modulus of the convolved complex signal was averaged over 0.1s fragments. Finally, every $i$-th window of the signal was represented with time-frequency representation in the form of a tensor $\underline{\mathbf{X}}_i \in \mathbb{R}^{64 \times 15 \times 10}$ with dimensions corresponding to ECoG channels, frequency bands, and time steps. 

In this study, samples for which predicted and desired states did not match each other were removed. By removing the gate errors, we minimize the influence of low gate model performance on the visual feedback and thus on the patient imagination patterns. In addition, one session was removed from the dataset as during the online experiment patient reached a highly negative cosine similarity (outliers compared to other sessions) which may as well influence recorded signals by providing erroneous visual feedback to the patient.

\subsection{UMAP embeddings and artifacts identification}
High-dimensional datasets are almost not possible to visualize without any dimensionality reduction before. What can be trivial to observe in low-dimensional space may easily stay hidden in the noise in high-dimensional representations. Due to the curse of dimensionality, understanding the topology of distributions or even noticing outliers is challenging. The main goal of the visualization was to see the evolution of data distributions between sessions. To map time-frequency representation into lower-dimensional space, an unsupervised learning algorithm, namely Uniform Manifold Approximation and Projection (UMAP) \cite{2018arXiv180203426M} was used. We decided to apply UMAP as it preserves the global manifold structure similarly to t-SNE \cite{kobak_initialization_2021} but has a lower computational load according to \cite{umap_performance,2018arXiv180203426M}. Thanks to that, we could avoid additional dimensionality reduction (e.g., PCA), which is usually done before feeding high-dimensional datasets into t-SNE \cite{vandermaaten08a}. We used flattened time-frequency features $\underline{\mathbf{X}}_i \in \mathbb{R}^{64 \times 15 \times 10} \rightarrow \mathbb{R}^{9600}$ (the same as for motor imagery decoding) as the input to UMAP. Every tenth observation from the dataset was selected for UMAP to avoid redundancy in the data (90\% overlap between samples) and decrease the computational load. UMAP was fitted on three datasets to all the sessions together, i.e., one UMAP for both hands optimized together and one per hand trained individually. The first scenario lets us better see the data distributions within the state classification framework, with samples being colored due to the state they belong to. This gave us a global overview of the dataset. In the per hand scenario, we focused more locally on the structure of each dataset. This may have a bigger influence on the decoding performance while being harder to analyze due to the lack of explicit labels for visualization (like states in the previous case).

In the case of UMAP optimized together for both hands, we proposed an indirect indicator of data quality reflecting the separability of the left and right hand clusters. This was assessed using linear support-vector machine (SVM). SVM was fitted to every session separately. Then every sample in the session was classified into two categories, i.e., left hand or right hand movement. Accuracy of the state classification was further used as a state separability indicator. We did not perform any cross-validation as we focused on the separability of the clusters and not on the state classification performance itself. On the UMAP embeddings we visualized also SVM decision boundary dividing the space between categories of movements.

\begin{figure}
	\centering \includegraphics[width=\columnwidth]{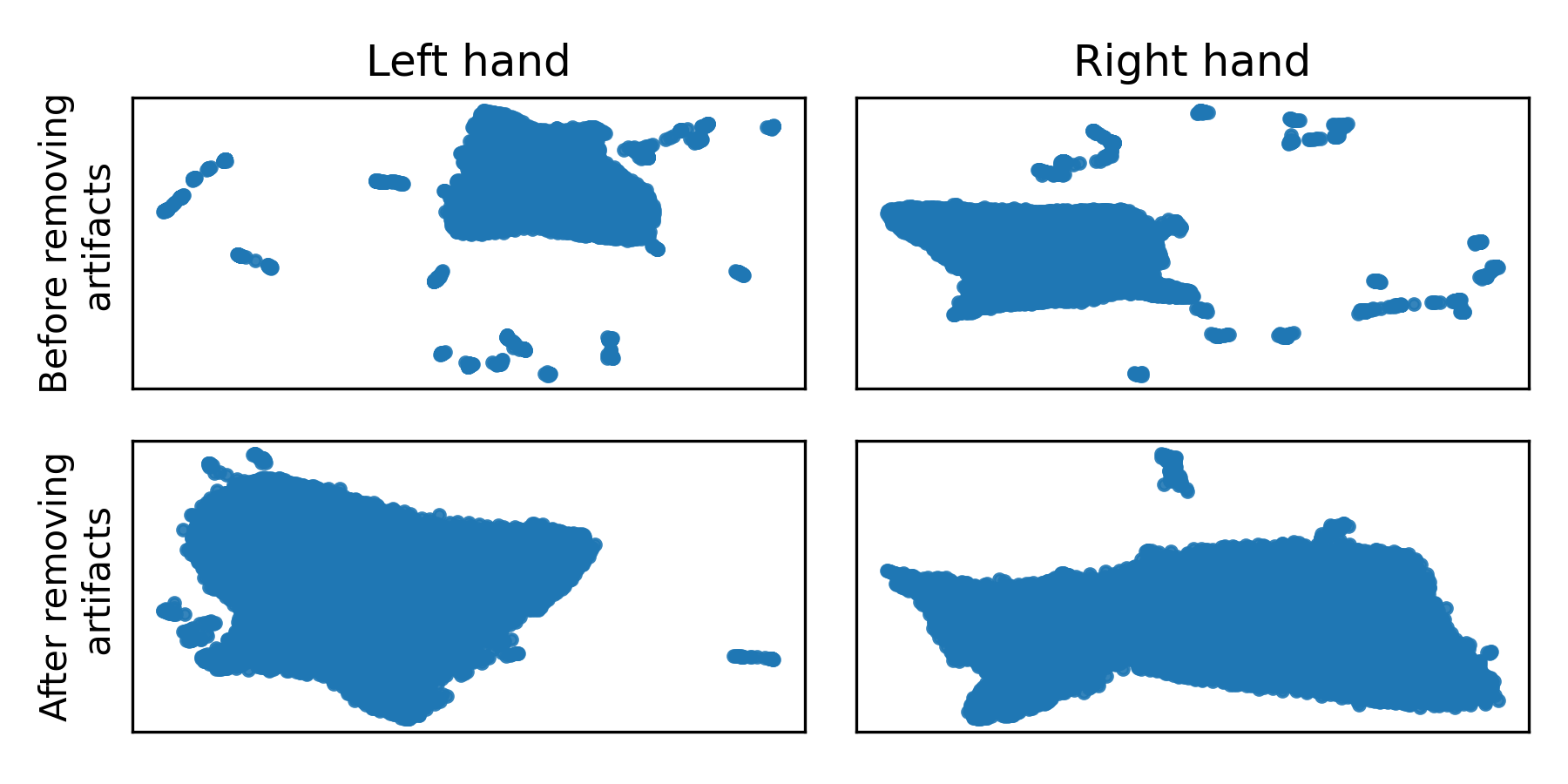}
	\caption{Per hand embeddings before (top row) and after (bottom row) artifacts removal.} \label{fig:umap-artifacts}
\end{figure}

UMAP as a dataset visualization method may also be used for an overall sanity check of the dataset, especially for artifacts that are easy to spot when the dataset is small, but it is impossible to review every sample individually when analyzing thousands of observations. In our case, UMAP helped us to observe artifacts coming from connection loss resulting in singular outliers samples that were not caught during recording. Those, on the UMAP plots, created suspicious clusters of observations (figure \ref{fig:umap-artifacts}). The clusters of artifacts after recognition on the UMAP plots and further manual review were fixed by interpolation of points in the raw signal domain.

\subsection{Evaluated models}
Multilinear model optimized with REW-NPLS algorithm \cite{eliseyev2017recursive} was used as a 'traditional' ML benchmark to predict 3D hand translation. The same algorithm was also used for providing online control to the patient during recordings. PLS models embed both high-dimensional input features and output variables into lower-dimensional latent space, aiming to extract latent variables with the highest correlation between input and output. REW-NPLS model can be updated online thanks to low-computational cost, recursive validation of the number of latent factors, and model parameters being updated with only chunks of the dataset. Online training ease performing the experiments and makes it possible to use ECoG decoders almost from the beginning of the first recording session. Even if decoders may show unstable performance at the beginning of the experiment due to the small amount of data used for training, it provides visual feedback to the patient. For our offline computational experiments, multilinear models were trained in pseudo-online mode, simulating real-life experiments with updates based on 15 seconds-long chunks of data.

The second group of models used deep learning to predict the desired hand translation. In particular, methods proposed and described in detail in \cite{sliwowski_2022} were evaluated---i.e., multilayer perceptron (MLP---simple approach) and mix of CNN and LSTM (CNN+LSTM+MT) providing the best performance for a given dataset \cite{sliwowski_2022}. MLP was built from two fully-connected layers with 50 neurons with dropout and batch normalization in-between (see table \ref{tab:mlp}). CNN-based method exploited the spatial correlation between electrodes by analyzing data organized on a grid reflecting the electrodes' arrangement with convolutional layers. As the CNN-based method utilizes data structure, it has fewer parameters while maintaining similar capabilities to MLP. In CNN+LSTM+MT, LSTMs were used to aggregate temporal information extracted by convolutional layers into desired translation trajectory (see table \ref{tab:cnn}). The DL models were trained to maximize cosine similarity (CS) between predicted and optimal trajectories. We used early stopping to limit the overfitting with a validation dataset consisting of the last 10\% of the calibration dataset. The best model on the validation dataset was used for further evaluations. The procedure was repeated five times for every scenario and model to limit the influence of the stochasticity of the training process on our results. To train DL models, we used a fixed set of hyperparameters, i.e., learning rate equals 0.001, weight decay (L2 regularization) equals 0.01, and batch size equals 200.

\begin{table}
\centering
\caption{MLP architecture from \cite{sliwowski_2022}.}
\label{tab:mlp}
\resizebox{\columnwidth}{!}{%
\begin{tabular}{lll} 
\toprule
Layer           & Kernel Shape & Output Shape   \\ 
\midrule
Flatten         & –            & {[}200, 9600]  \\
Fully connected & {[}9600, 50] & {[}200, 50]    \\
BatchNorm       & {[}50]       & {[}200, 50]    \\
ReLU            & –            & {[}200, 50]    \\
Dropout         & –            & {[}200, 50]    \\
Fully connected & {[}50, 50]   & {[}200, 50]    \\
BatchNorm       & {[}50]       & {[}200, 50]    \\
ReLU            & –            & {[}200, 50]    \\
Dropout         & –            & {[}200, 50]    \\
Fully connected & {[}50, 3]    & {[}200, 3]     \\
\bottomrule
\end{tabular}
}
\end{table}

\begin{table}
\centering
\caption{CNN+LSTM+MT architecture from \cite{sliwowski_2022}.}
\label{tab:cnn}
\resizebox{\columnwidth}{!}{%
\begin{tabular}{lll} 
\toprule
Layer                         & Kernel Shape        & Output Shape           \\ 
\midrule
Input                         &                     & {[}200, 15, 8, 8, 10]  \\
\hspace{5mm}Input per implant &                     & {[}200, 15, 8, 4, 10]  \\
\hspace{5mm}Conv space        & {[}15, 32, 3, 3, 1] & {[}200, 32, 6, 4, 10]  \\
\hspace{5mm}ReLU              & –                   & {[}200, 32, 6, 4, 10]  \\
\hspace{5mm}BatchNorm         & {[}32]              & {[}200, 32, 6, 4, 10]  \\
\hspace{5mm}Dropout           & –                   & {[}200, 32, 6, 4, 10]  \\
\hspace{5mm}Conv space        & {[}32, 64, 3, 3, 1] & {[}200, 64, 4, 2, 10]  \\
\hspace{5mm}ReLU              & –                   & {[}200, 64, 4, 2, 10]  \\
\hspace{5mm}Dropout           & –                   & {[}200, 64, 4, 2, 10]  \\
LSTM                          & –                   & {[}200, 10, 50]        \\
LSTM                          & –                   & {[}200, 10, 3]         \\
\bottomrule
\end{tabular}
}
\end{table}

\subsection{Computational experiments}
Multiple offline computational experiments were performed on the prerecorded ECoG BCI dataset to assess the impact of training dataset size on decoding performance. The results computed on a real-life dataset may be impacted by multiple factors that cannot be observed directly. Thus, we proposed several ways of increasing the training dataset as well as iterating over it. By modifying the training datasets in different manners, we aimed to isolate different factors that can potentially influence learning curves. In every scenario, all the models were trained on a different subset of the database and then evaluated on test datasets accordingly to the experiment.

\subsubsection{Forward increase}
The forward increase (FI) scenario measured the change of cosine similarity when adding more recording sessions to the dataset. This experiment corresponds to a real-life situation where more data is collected during the experiment. The sessions were incrementally added (session by session) to the training dataset. After every step, all the decoders were trained from scratch and evaluated on the following 22 sessions (see Figure \ref{fig:experiments}).

\subsubsection{Backward increase}
An important factor influencing model training may be the nonstationarity of signal in time originating from the plasticity of the brain as well as the patient's adaptation. To assess the influence of this factor, an inverse of forward increase was performed, further referred to as backward increase (BI). Similar to the FI simulation, the training dataset was increased session by session. However, the increase was started from the 21st session and the previous sessions were added until including the first calibration session. After every training, models were evaluated on a fixed test set consisting of 22 last recordings (see Figure \ref{fig:experiments}).

\subsubsection{Random increase}
An alternative way of assessing the influence of training dataset size on the decoder performance is random dataset increase (RI). Instead of maintaining the temporal order of recorded samples, we artificially removed the connection between neighboring observations, i.e., for every dataset size, a respective number of observations was selected from the first 22 sessions, and then the model was trained. This may reduce the effects of neuronal signal nonstationarity and/or patient adaptation and provide results closer to theoretical learning curves when assumptions about the stationarity of observations are fulfilled. Evaluations were performed on the same test set as in BI.

\subsubsection{Dataset translation}
\begin{figure}
	\centering \includegraphics[width=\columnwidth]{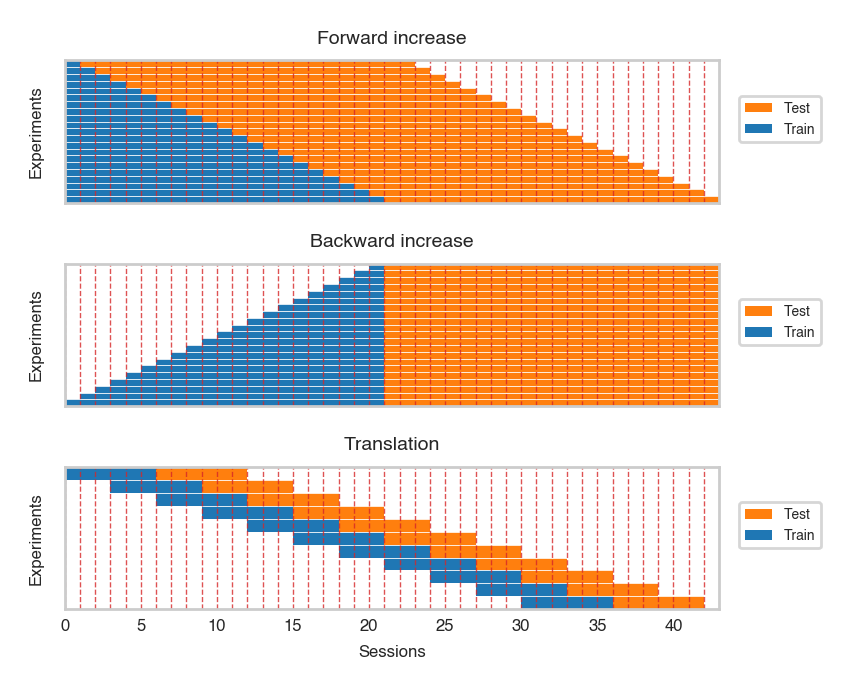}
	\caption{Visualization of forward and backward increase and translation over the dataset. For clarity, we ignored differences in session length.} \label{fig:experiments}
\end{figure}

As data may change over time, we trained models on approximately the same amount of data but recorded in different periods of the experiment. This enabled us to rule out the effect of the increased dataset and focus on data shift and potential patient adaptation that may modify the data representation and influence the performance of trained decoders. The training dataset was translated over the whole dataset and evaluated on the test dataset consisting of the following six sessions (see figure \ref{fig:experiments}).

\subsection{Learning curve}
The learning curve \footnote{In this context, the learning curve does not refer to the relationship between the number of training epochs and model performance which the name learning curve is also commonly used for.} describes the relationship between model performance and the training dataset size $l$ \cite{NIPS1993_power_law}. It can be used, for example, to infer a potential change in the performance from adding more data to the system. This can be particularly efficient in application to BCI because we can estimate the hypothetical performance of decoders when recording more data without actually performing the experiments. Learning curves can also be used to select an appropriate model for a specific dataset size. For example, Strang et al. \cite{strang-linear} showed that non-linear models are more likely to outperform linear models for bigger datasets. On the other hand, Hoiem et al. \cite{Hoiem2021LearningCF} showed that models with more parameters can be more efficient in the case of small datasets despite the higher potential for overfitting.

The learning curve may be formulated with power law \cite{NIPS1993_power_law, baouhua_learning_curve}. In our case, the relationship between cosine similarity and training dataset size may be expressed as:
\begin{equation}
    \mathrm{CS}(l; a, b, c) = a - b \cdot l^{-c}
\end{equation}
where $b$ and $c$ can be interpreted as learning rate and decay rate, respectively. $a$ corresponds to theoretical asymptotic performance when $l\to\infty$. Parameters $a$, $b$, and $c$ were fitted to the results obtained in RI experiment with non-linear least squares using Trust Region Reflective algorithm with bounds $a \in [-1, 1]$, $b > 0$, and $c > 0$.

\subsection{Intrinsic dimensionality estimation}
The idea of patient adaptation and improving BCI skills using visual feedback is based on the assumption that the patient can modify/adjust motor imagery patterns to solve the task better. As a result, the data distribution and the shape of the data manifold may change. To estimate the data distribution changes, intrinsic dimensionality (ID) estimation methods may be used. ID reflects the minimum number of variables needed to represent the dataset without a significant information loss. Thus, the ID indicator is strictly connected to a dataset's true dimensionality, which is an important factor in data analysis, influencing the performance and changing the number of samples needed to train models. Intuitively, in a typical case, higher-dimensional manifolds are harder to learn due to the 'curse of dimensionality.' ID is better studied for images that, although have thousands of pixels, lie on a lower-dimensional manifold (e.g., less than 50 for ImageNet \cite{pope2021the}). We use ID as a potential data quality indicator, which may vary with different recording sessions. ID estimates were computed for every session, and values from the respective sessions were averaged to obtain training dataset estimates for the dataset translation experiment.
% on the datasets used to train models in the dataset translation experiment. 
\begin{figure*}
	\centering \includegraphics[width=\textwidth]{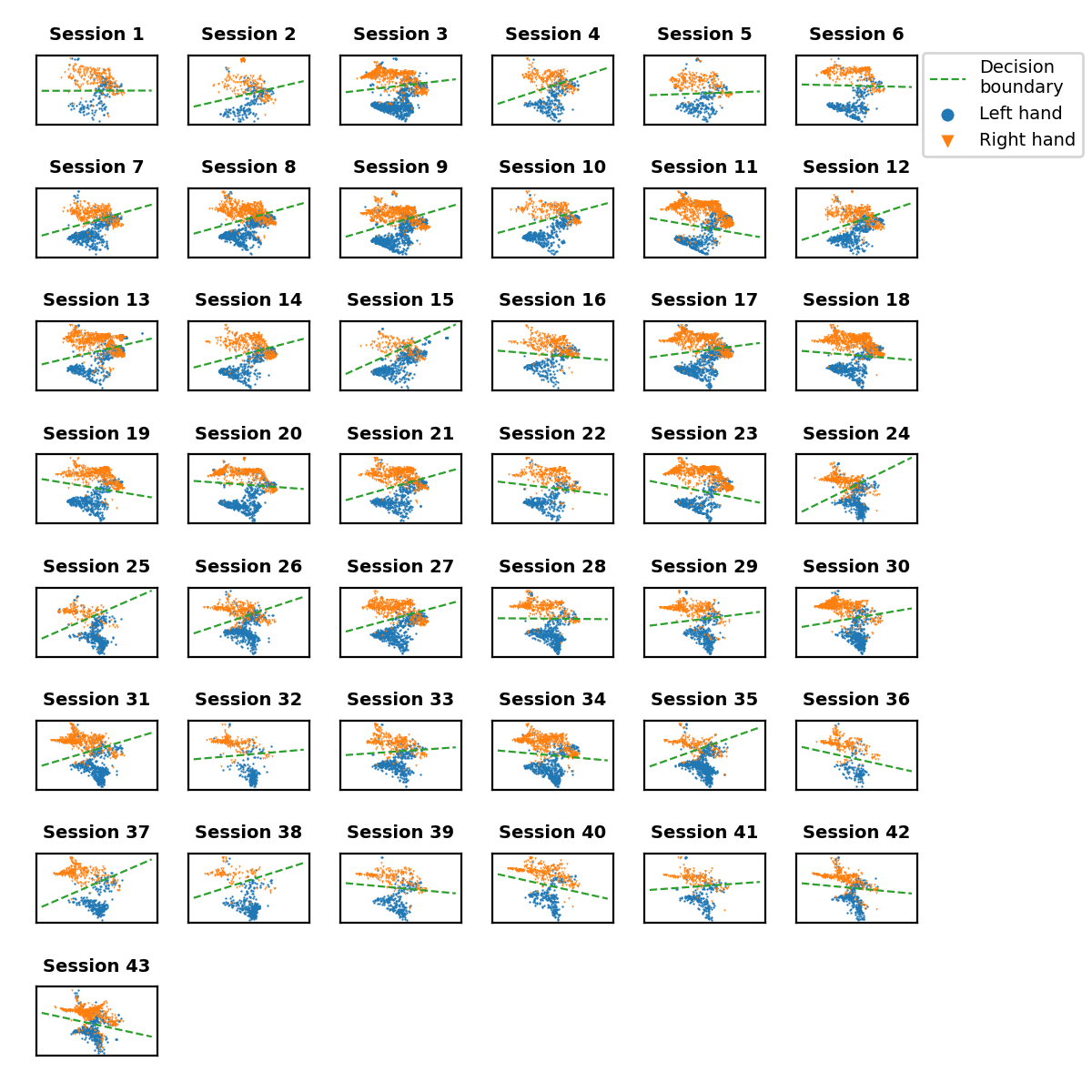}
	\caption{Visualization of 2D embedding of left and right hand data obtained using UMAP. Green dashed line showes SVM decision boundary.}
	\label{fig:embedding-per-session}
\end{figure*}
To compute ID, we used current state-of-the-art methods, namely expected simplex skewness (ESS) \cite{ess_2015} estimating local ID in data neighborhoods (in our case 100 points) and TwoNN \cite{Facco2017EstimatingTI} estimating global dataset ID. ESS, according to \cite{tempczyk_lidl} provides better estimates for high ID values, while most of the methods tend to underestimate the ID (e.g., TwoNN \cite{Facco2017EstimatingTI}). It is especially important because our preliminary analysis showed that ECoG data is high dimensional, with ECoG features' mean local ID being significantly higher than the mean local ID for images (around 300 for ECoG, below 15 for MNIST, EMNIST, FMNIST \cite{bac_lid_concentration}). For ID computations we used scikit-dimensions package \cite{scikit-dimensions}.
% We did not use LIDL \cite{tempczyk_lidl} due to the high computational load in the preliminary analysis.

\section{Results}

\subsection{UMAP}

\begin{figure}
	\centering \includegraphics[width=.8\columnwidth]{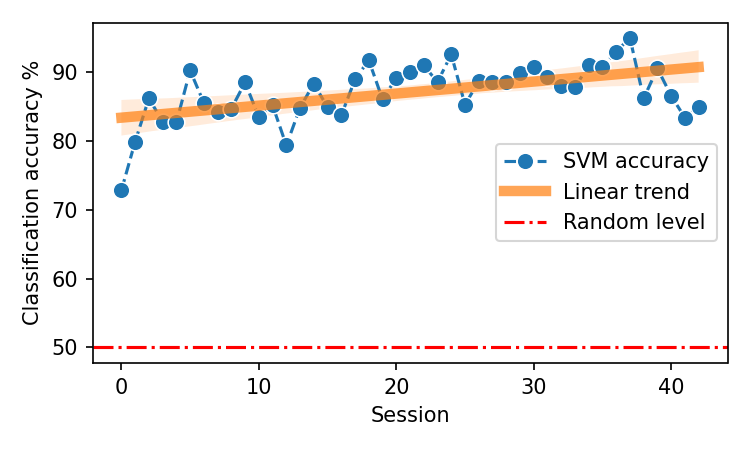}
	\caption{Accuracy of left vs. right hand state classification using SVM classifier. The orange line indicates a linear trend fitted to the points.}
	\label{fig:svm-accuracy}
\end{figure}

Data distributions for every session were shown in figure \ref{fig:embedding-per-session} with colors indicating left and right hand states. With time, clusters of states get better separated from each other. We quantified separability of different states with SVM classification accuracy (figure \ref{fig:svm-accuracy}). An increase in accuracy can be observed for sessions recorded later in the experiment, with a maximum accuracy of 95\% for session 37. Note that UMAP, similarly to t-SNE, does not preserve the density of points when mapping to the lower dimensional space and may, in some cases, create sub-clusters that originally may not exist in the input space.

\subsection{Forward and backward increase}
\begin{figure*}
	\centering \includegraphics[width=\textwidth]{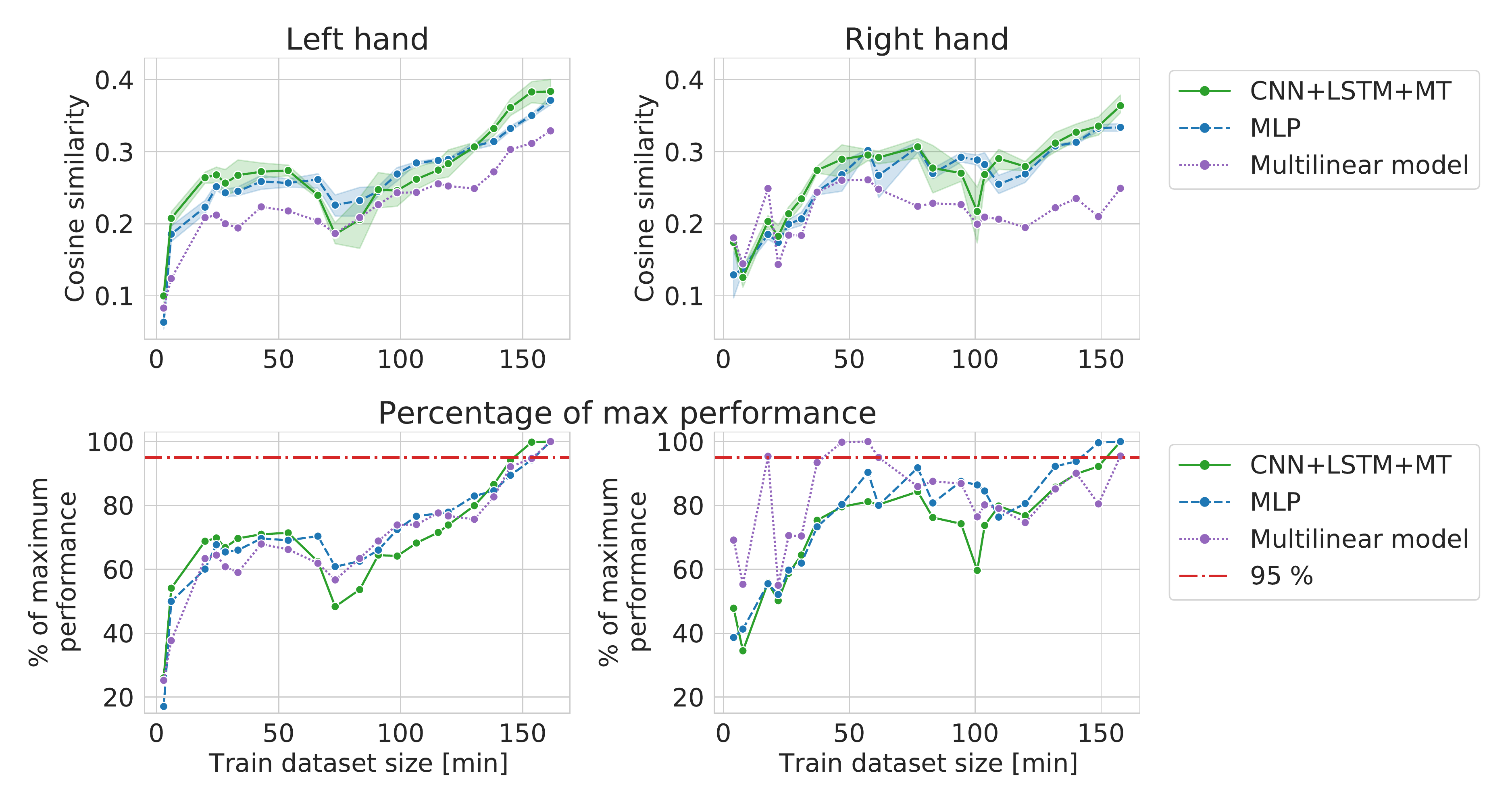}
	\caption{Cosine similarity computed in forward increase experiment, i.e., different training dataset sizes when starting from the first session.} \label{fig:forward-increase}
\end{figure*}

Forward increase results (figure \ref{fig:forward-increase}) show learning curves in a situation close to a real-life scenario when more recordings are performed in the experiment. For all the models, a sharp increase in performance can be observed for small datasets. After 30-40 minutes of data, the curves become flat, reaching 70-80\% of maximum FI performance (except 100\% for the multilinear right-hand model) until 100-120 minutes of the signal. For datasets with more data than 100-120 minutes, a slow performance increase can be noticed. In the case of the left hand dataset, it starts earlier and is also visible for the multilinear model, while for the right hand, REW-NPLS performance stays stable. Overall, multilinear and DL models have similar learning curves and reach a performance plateau after including the same amount of data. However, multilinear models usually perform worse than DL models for the same amount of data.

\begin{figure*}
	\centering \includegraphics[width=\textwidth]{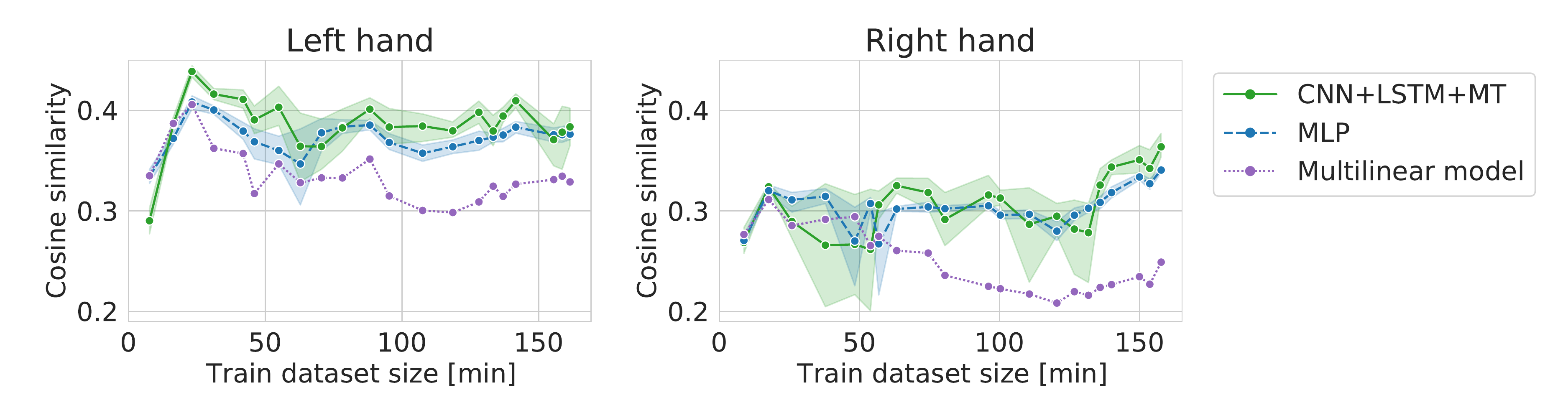}
	\caption{Cosine similarity for backward increase experiment, i.e., different training dataset sizes when starting from the 21st session and going backward.} \label{fig:backward-increase}
\end{figure*}

Extending the dataset backward, starting from the middle of the recorded dataset, does not correspond to any real-life scenario. However, by doing this, we were able to assess the potential influence of data quality change on the results computed in the FI computational experiment. In the case of backward increase (figure \ref{fig:backward-increase}), high performance can be observed for relatively small datasets---with just 3 (left hand) and 2 (right hand) sessions. For bigger datasets, the performance stabilizes or slightly decreases. The curves for all the models behave similarly. Performance of DL models starts to increase for $>130$ minutes of signal for the right hand and achieves the best cosine similarity. When comparing FI and BI, in the case of the left hand, the best performance can be observed for BI and only 3 sessions of data in the training dataset. In the case of the right hand, the highest performance is achieved for the biggest dataset, suggesting that recording more data may improve the cosine similarity. The small amount of data needed to achieve high performance (2-3 sessions) in the BI experiment may suggest neuronal patterns improvement resulting in dataset quality increase (the amount of data required to reach a given performance).

\subsection{Random increase}
In the RI experiment, the influence of patient adaptation and signal nonstationarity is reduced as all the links between neighboring samples are destroyed when selecting data for the training dataset. Results for RI are more similar to theoretical learning curves of DL models, with a sharp increase in performance in the beginning and saturation when the model's maximum capacity is achieved. The performance is saturated after adding approximately 60-90 minutes of data to the training dataset at 95\% of maximum cosine similarity for RI experiment. Only a small improvement can be observed from using more data. For the multilinear model, we can observe that saturated best performance is lower than in the case of DL models. DL methods are able to learn more complex functions and thus can reach higher performance.
% When comparing DL models, MLP shows slightly higher performance with more stable cosine similarity, especially for the right hand dataset.
Fitted learning curves show the relationship between cosine similarity and dataset size within a theoretical framework, emphasizing the bigger capabilities of DL methods. The best models trained in the RI experiment showed lower performance compared to the best models from other experiments (dataset translation for both hands and BI for the left hand). However, in every experiment except BI and RI models were evaluated on different test datasets (see figure \ref{fig:experiments}). 

\begin{figure*}
    \centering
    \includegraphics[width=\textwidth]{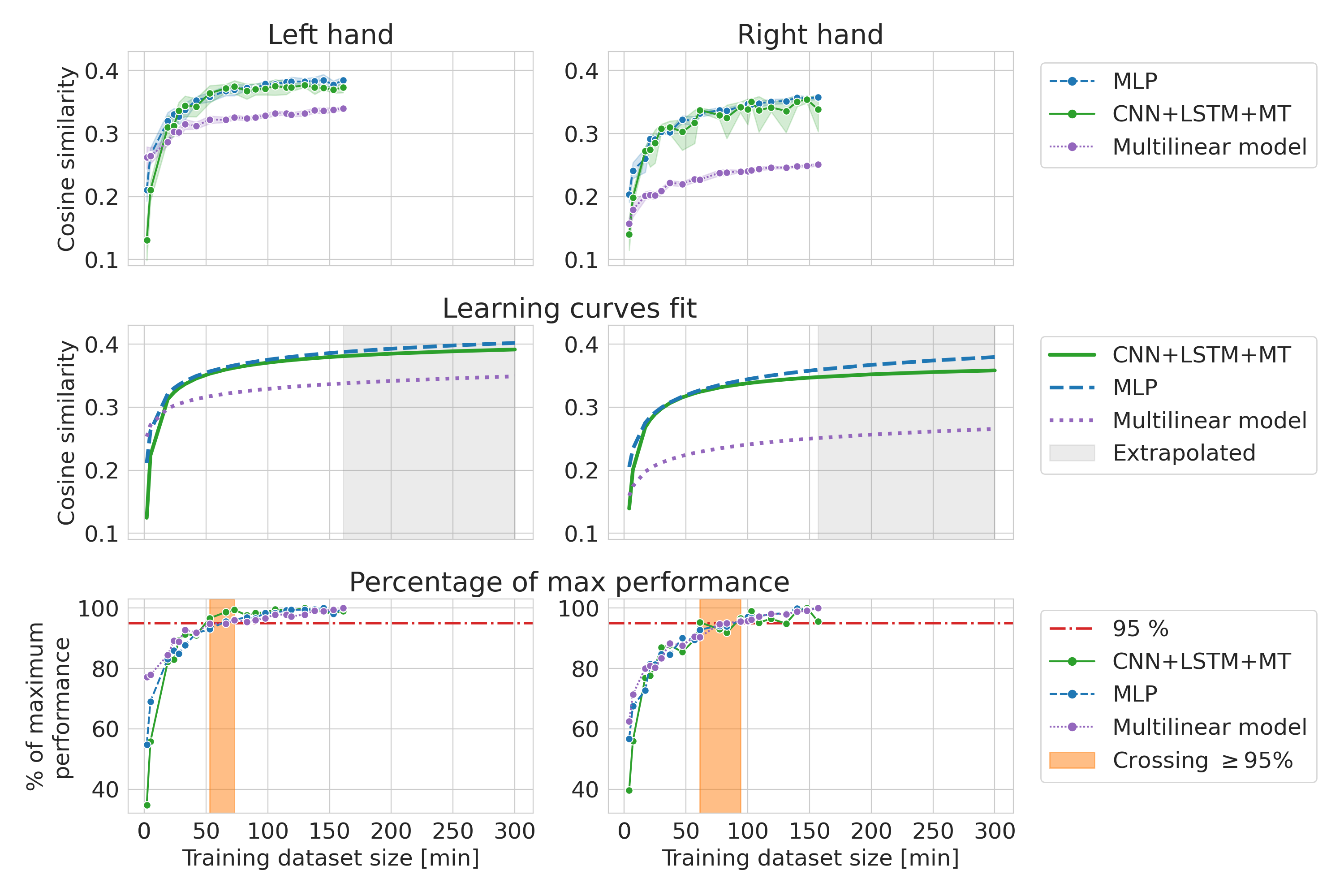}
    \caption{Cosine similarity for random increase experiment, i.e., different training dataset sizes when randomly selecting a subset of observations from the first 22 sessions. Every evaluation was performed 10 times.}
    \label{fig:random_increase}
\end{figure*}

\subsection{Dataset translation}
The dataset translation experiment shows the change in performance while maintaining approximately the same amount of data (six sessions) in the training dataset. Generally, all models show similar trends. For the left hand, we can observe an increase in cosine similarity for datasets recorded later in the experiment suggesting an improvement in data quality. The increase is less visible for the right hand dataset. This is confirmed by the slope of the linear trend fitted to the average performance of all the models (table \ref{tab:dataset-translation-params}). Expected cosine similarity improvement from training a model on the dataset recorded later was equal to 0.0069 per session and 0.0044 per session for left and right hand datasets, respectively. For both datasets, the most significant performance increase between the first and last evaluation can be observed for the multilinear model (table \ref{tab:diff-dataset-translation}). It may suggest that patient, to some extent, adapted specifically to the linear model family. The multilinear model does not follow the same fluctuations as the DL methods. The difference could be caused by the way of validating models (10\% validation set for DL, recursive validation on last 15 seconds of data at every step for pseudo-online REW-NPLS).
\begin{figure*}
	\centering \includegraphics[width=\textwidth]{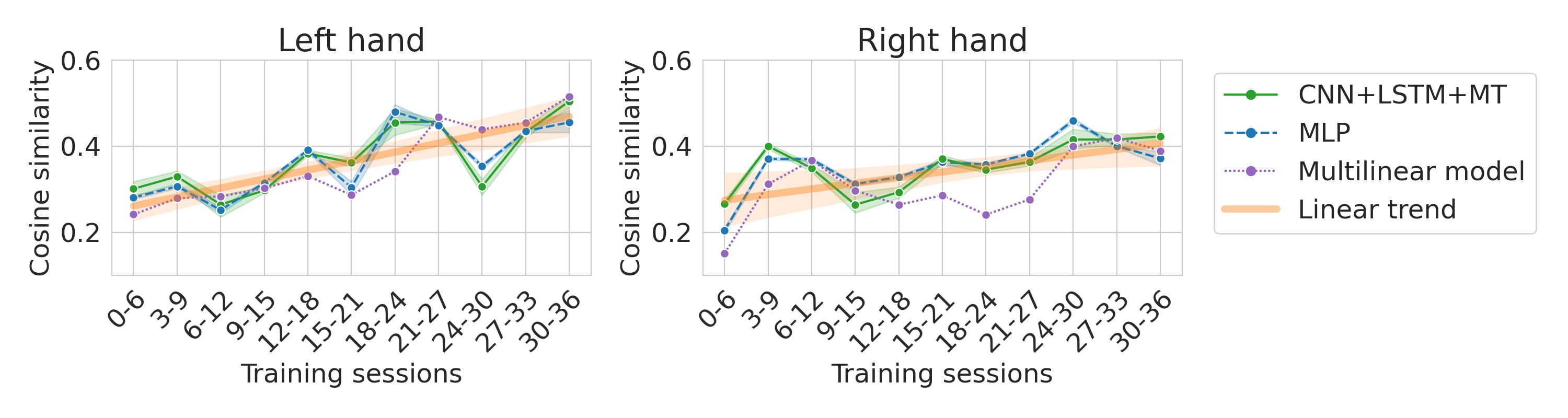}
	\caption{Cosine similarity for dataset translation, i.e., different training datasets (always 6 sessions for training and following 6 sessions for testing) translated over the dataset. The orange line indicates a linear trend line fitted to the models' average.} \label{fig:6on6-translation}
\end{figure*}

\begin{table}
\centering
\caption{Differences between models trained on sessions 0-6 and 30-36 in the dataset translation experiment.}
\label{tab:diff-dataset-translation}
\begin{tabular}{lrr}
\toprule
 &  Left hand &  Right hand \\
\midrule
CNN+LSTM+MT       &   0.203 &    0.157 \\
MLP               &   0.175 &    0.167 \\
Multilinear model &   0.274 &    0.239 \\
\bottomrule
\end{tabular}
\end{table}

\begin{table}
\centering
\caption{Parameters of trend lines fitted to the dataset translation results. Statistically significant p-values for the correlation coefficient are marked with asterisks.}
\label{tab:dataset-translation-params}
\begin{tabular}{lcccc}
\toprule
{} &   Slope &  Intercept &       R &  p-value \\
\midrule
Left hand  &  0.0069 &     0.2612 &  0.8816 &   0.0003* \\
Right hand &  0.0044 &     0.2744 &  0.6999 &   0.0165* \\
\bottomrule
\end{tabular}
\end{table}

\begin{figure*}
	\centering \includegraphics[width=\textwidth]{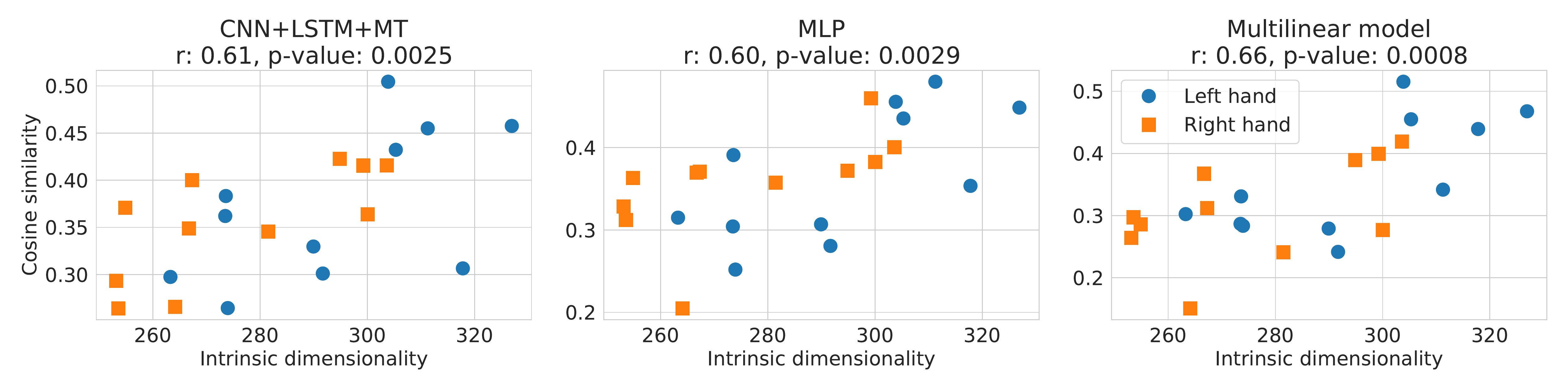}
	\caption{Relationship between cosine similarity and local ID of the training dataset computed with ESS for dataset translation experiment. In the plot titles, Pearson correlation coefficient r and p-value (the probability of two uncorrelated inputs obtaining r at least as extreme as obtained in this case) are presented.} \label{fig:lid-vs-cs}
\end{figure*}

In figure \ref{fig:lid-vs-cs}, the relationship between the local ID of the training dataset computed with ESS and the cosine similarity of different models for the translation experiment is presented. A statistically significant ($\alpha < 0.05$) correlation between local ID and models' performance was observed for all the methods, reaching up to 0.66 of the r correlation coefficient for the multilinear model. An overall trend of achieving higher cosine similarity can be observed for training datasets with a higher ID. ID for the analyzed datasets varies between 250 and 330, which is much more compared to less than 15 reported for MNIST, EMNIST, FMNIST \cite{bac_lid_concentration}.

\section{Discussion}

% \subsection{Training dataset size influence}
Our results showed that including more data in the training dataset for ECoG BCI may not be immediately visible on the performance metrics if already having access to 40 minutes of the signal. Indeed, a drastic increase in performance can be noticed for datasets smaller than 40 minutes. This justifies the current experimental paradigms in which 40-50 minutes of the signal is collected (corresponding to achieving approximately 70-80\% of maximum performance achieved with datasets up to 160 minutes of data) for training 3D hand translation models.

Theoretically, models with bigger capacity can benefit stronger from having access to more data. One of the indicators of model capacity can be the number of trainable parameters. In our case, MLP had the biggest number of trainable parameters (482 953), followed by CNN+LSTM+MT (238 772) and the multilinear model (28 800). The difference in potential performance gains can be visible in figure \ref{fig:random_increase}. For small datasets, a multilinear model outperforms DL-based approaches (left hand) or provides approximately the same cosine similarity. However, the multilinear model saturates at a lower level of cosine similarity, resulting in a performance gap which could be explained by the difference in model capacity. Multilinear models are more likely to provide high performance compared to DL for small datasets, which is consistent with the ML theory of less complex functions being less prone to overfitting. RI results and fitted theoretical learning curves revealed models' characteristics while limiting the influence of other factors like distribution shifts or patient adaptation on the decoding performance. Finally, all models saturate for relatively small training datasets (50-90 minutes for RI, 50 minutes for FI, 30 minutes for BI) with only slight improvement from adding more data ($\sim 5$\%). This amount of data is similar to the usual amount of data used in BCI studies. 

While this result validates previously developed data processing and experimental pipelines, a question arises whether it is an actual property/characteristic of brain signals or the shape of the curve is influenced by the previous years of research in which a relatively small amount of data was usually used to develop pipelines. There are hundreds of hyperparameters influencing data processing characteristics, starting from recording devices (e.g., number of electrodes, mental task design), signal processing pipelines (e.g., electrodes montage, filtering, standardization), ending on hyperparameters of machine learning models of all kinds (e.g., models' capacity, regularization weight, the architecture of models). The lack of huge improvement from increasing the dataset may be caused just because we reached the level of decoding close to maximum due to a lack of information in the data needed for prediction. However, from another perspective, one can hypothesize that we observe an effect of researchers overfitting to the specific conditions observed so far.

\subsection{Models optimization for big datasets}
All offline experiments were performed with a fixed set of hyperparameters. At the same time, different dataset sizes may require a change in the hyperparameters. For example, regularization limits overfitting, which should be less severe when the training dataset is big. Similar logic applies to dropout, which limits overfitting but on the other hand, it decreases models' capacity by introducing redundancy in the network representation. In the BI experiment, we observed a decrease in performance when adding more data for the left hand dataset. Hypothetically, increasing models' capacity may solve this problem (assuming it is caused by adding samples from different distributions to the dataset) because models with bigger capacity might not have to 'choose' on which motor imagery patterns they should focus. However, hyperparameters search is time and resource-consuming, so performing hyperparameters search for every dataset size may not be reasonable. In the future, DL architectures with bigger capacities in terms of the number of layers, number of neurons, etc., should be evaluated. 

Datasets can also be artificially extended by using data augmentation methods. A variety of beneficial data augmentation methods exist for brain signals, especially EEG \cite{Rommel2022DataAF}, that might improve decoding accuracy for the 3D hand movement control. Hoiem et al. \cite{Hoiem2021LearningCF} showed, for computer vision datasets, that data augmentation may act as a multiplier of the number of examples used for training. In the light of recent advancements in EEG data augmentation, i.e., class-wise automatic differentiable data augmentation \cite{rommel2022cadda}, it can be interesting to investigate how the reported results generalize to ECoG signals and influence, presented here, learning curves.

\subsection{Patient training}
UMAP embeddings may reveal interesting data manifold structures. In our case, we observed signs of distribution change on the embeddings visualization and the separability of left/right hand observations. Points start to be distributed denser in some regions of the plots and align along lines (see for examples sessions 35, 42, 43 for the left hand in figure \ref{fig:embedding-lh} or sessions 31, 41, 42 in figure \ref{fig:embedding-rh}). Additionally, in the dataset translation experiment, we can see an increase in cosine similarity, stronger for the left hand. Moreover, the overall best performance for the left hand was achieved with only 3 sessions ($\sim 25$ minutes of signal), outperforming models trained on much bigger datasets. This suggests improvements in patient BCI skills by adapting motor imagery patterns to the ML pipeline used in the study but non-specific to the multilinear model because trends are visible for all the evaluated approaches. At the same time, adding more data with noisy and changing patterns may not be profitable for the predictions. Thus, more focus should be placed on obtaining high-quality and well-separable motor imagery patterns in the signal. Patient adaptation is possible thanks to the visual feedback provided to the participant during recordings. The potential for patient adaptation creates a perspective for further improvements of BCI performance with the experience gained by the patient in long-term usage of the system. However, the reason adaptation is visible only for the left hand remains unknown. We hypothesize that the motor imagery patterns are easier to adapt for the left hand thanks to the remaining residual control resulting in better cortex preservation. 

Our results showed a correlation between the local ID of the training dataset and the models' performance. This may indicate that models achieve better results when trained on more complicated manifolds. However, this hypothesis is counterintuitive and contradictory to research in computer vision. Thus, we hypothesize that higher ID may also indicate more diverse motor imagery patterns, better representing those found in the test set. Diversity of patterns may be harmful to models with a too-small capacity to learn them all. However, to some extent, it may be helpful as it creates a more diverse dataset that better reflect/cover the real manifold of all motor imagery patterns. Finally, we hypothesize that another hidden factor affects both the local ID and the amount of information needed for prediction, like the diversity of motor imagery patterns, so a change in local ID may not cause the increase in the performance itself. For example, local ID can also be increased by adding Gaussian noise to the signal, decreasing cosine similarity instead. Investigation of this kind of relationship is especially challenging in the case of brain signals due to a lack of data understanding with the 'naked eye,' which would significantly ease finding a correct interpretation of observed phenomena. As a next step, more ID estimation methods could be evaluated as statistically significant correlations for DL models were observed only for local ID computed with ESS. In the case of TwoNN, global ID did not show a significant correlation for DL approaches (see figure \ref{fig:lid-vs-cs-twonn}). This could be caused by worse TwoNN precision for high ID values as well as a lack of local per-sample ID information in the global ID dataset estimate. The relationship between local ID and performance should be further analyzed on different brain signal datasets.

% For instance, DL models with a large number of layers have been shown to be less performant than shallow architectures on relatively small dataset \cite{schirrmeister2017deep}.

\subsection{Interpretation limitations}
All the computational experiments analyzed in this study were obtained offline using data recorded with only one patient. Thus the learning curves and potential of patient adaptation should be further investigated in a bigger population with online experiments verifying our conclusions. Specifically, an online experimental protocol aiming to isolate patient training (with or without visual feedback) and decoders update influence on performance should be designed.

Our results were computed on a real-life dataset recorded with a tetraplegic patient. Analyzing this kind of dataset allows us to draw conclusions about the population in real need of assistive technology. However, interpretation of results is even more challenging than in the case of healthy subjects because we do not have access to solid ground-truth labels to train ML models. This increases the already long list of factors that can affect the performance of the decoders and may not be easily noticed when analyzing ECoG signals. For example, in the ideal ML world, one could analyze the learning curve and draw conclusions about the required dataset size to effectively train ML models. In our case, other factors like the nonstationarity of the signal play an important role in the process. In some cases, we may add more data to the dataset (e.g., BI experiment) and decrease the performance because we also 'extend' the data manifold with samples from a shifted distribution. A remedy for distribution shifts between sessions may be methods used for domain adaptation, for example, in EEG transfer learning \cite{domain_adaptation}. Part of the aforementioned issues limiting our interpretation capabilities might be addressed with generative models \cite{goodfellow_gan} that are a popular tool in computer vision. In the case of brain signals, the ability to produce signals with the given parameters and characteristics may be used to verify and understand phenomena observed in real-life experiments. First attempts to train GANs for EEG \cite{hartman_eeg_gan} data analysis were made, but a significant amount of work still has to be done to create a consistent framework for easier hypothesis evaluation.

\section{Conclusions}
Our analysis showed that adding more data to the training dataset may not be instantly profitable, starting from datasets containing 30-50 minutes of the signal in real-life scenarios. Instead, improvement may be achieved by creating a high-quality dataset that can be recorded after participant training. Furthermore, we showed the importance of patient adaptation in the human-in-the-loop system that enabled obtaining high-performance models with relatively small training datasets. Finally, we propose UMAP embeddings and local intrinsic dimensionality as a way to visualize the data and potentially evaluate data quality.

% \begin{itemize}
%     \item Extending training dataset may not be profitable in short term perspective (starting from 40 minutes of data)
%     \item Data quality affects the performance drastically
%     \item Data quality increases with time - patient's adaptation
%     \item Intrinsic dimensionality may be an indicator of data quality which correlates with cosine similarity
%     \item UMAP can be used for artifacts cleaning and distribution trends visualization
    
% \end{itemize}

\section*{Data availability statement}
The data analyzed during the current study are not publicly available for legal/ethical reasons but are available from the corresponding author on reasonable request.

\ack{
Clinatec is a Laboratory of CEA-Leti at Grenoble and has statutory links with the University Hospital of Grenoble (CHUGA) and with University Grenoble Alpes (UGA). This study was funded by CEA (recurrent funding) and the French Ministry of Health (Grant PHRC-15-15-0124), Institut Carnot, Fonds de Dotation Clinatec. Matthieu Martin was supported by the cross-disciplinary program on Numerical Simulation of CEA. Maciej Śliwowski was supported by the CEA NUMERICS program, which has received funding from European Union's Horizon 2020 research and innovation program under the Marie Sklodowska-Curie grant agreement No 800945.}

\appendix

\section{UMAP embeddings}

\begin{figure*}
	\centering \includegraphics[width=\textwidth]{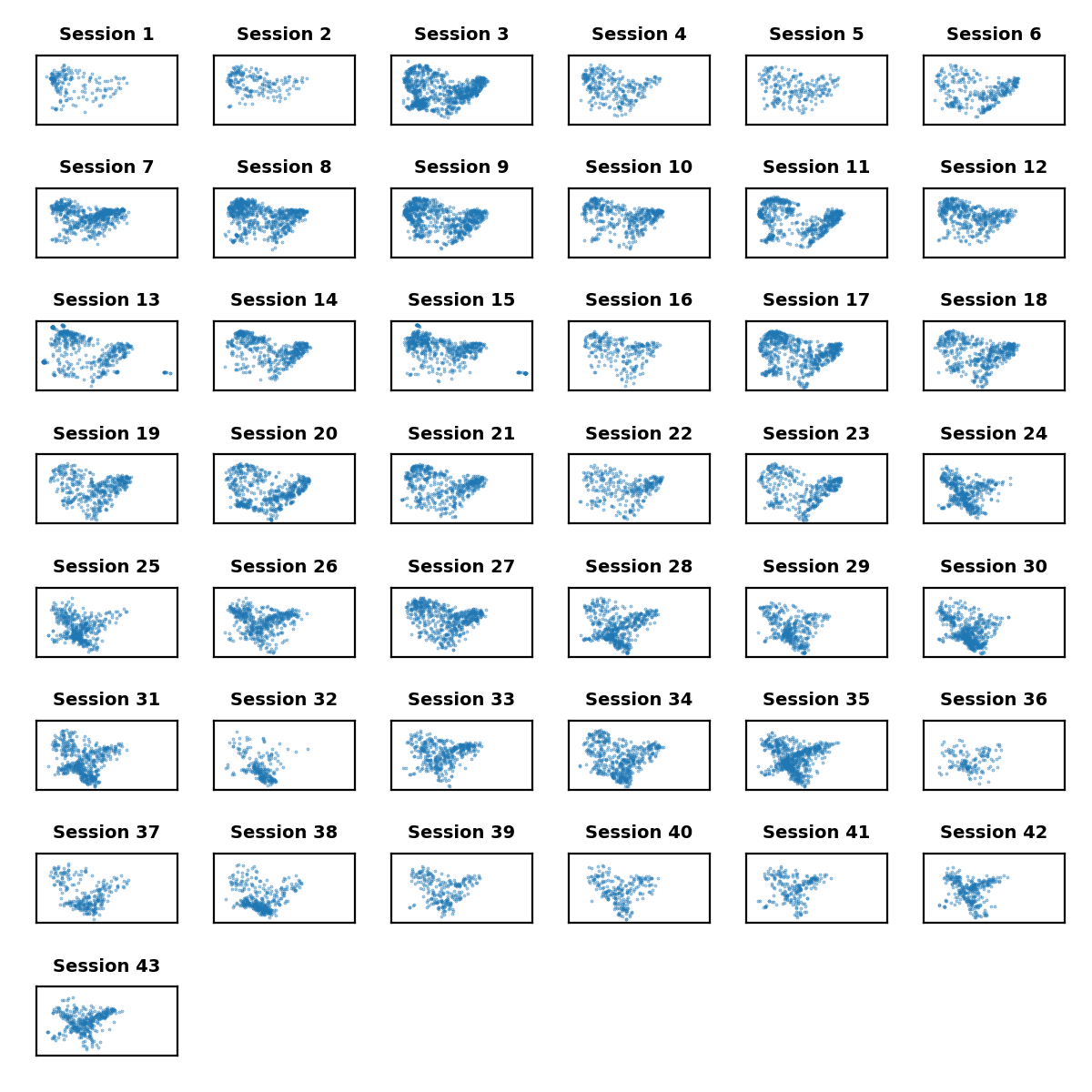}
	\caption{Left hand dataset embedding showed for every session separately.} \label{fig:embedding-lh}
\end{figure*}

\begin{figure*}
	\centering \includegraphics[width=\textwidth]{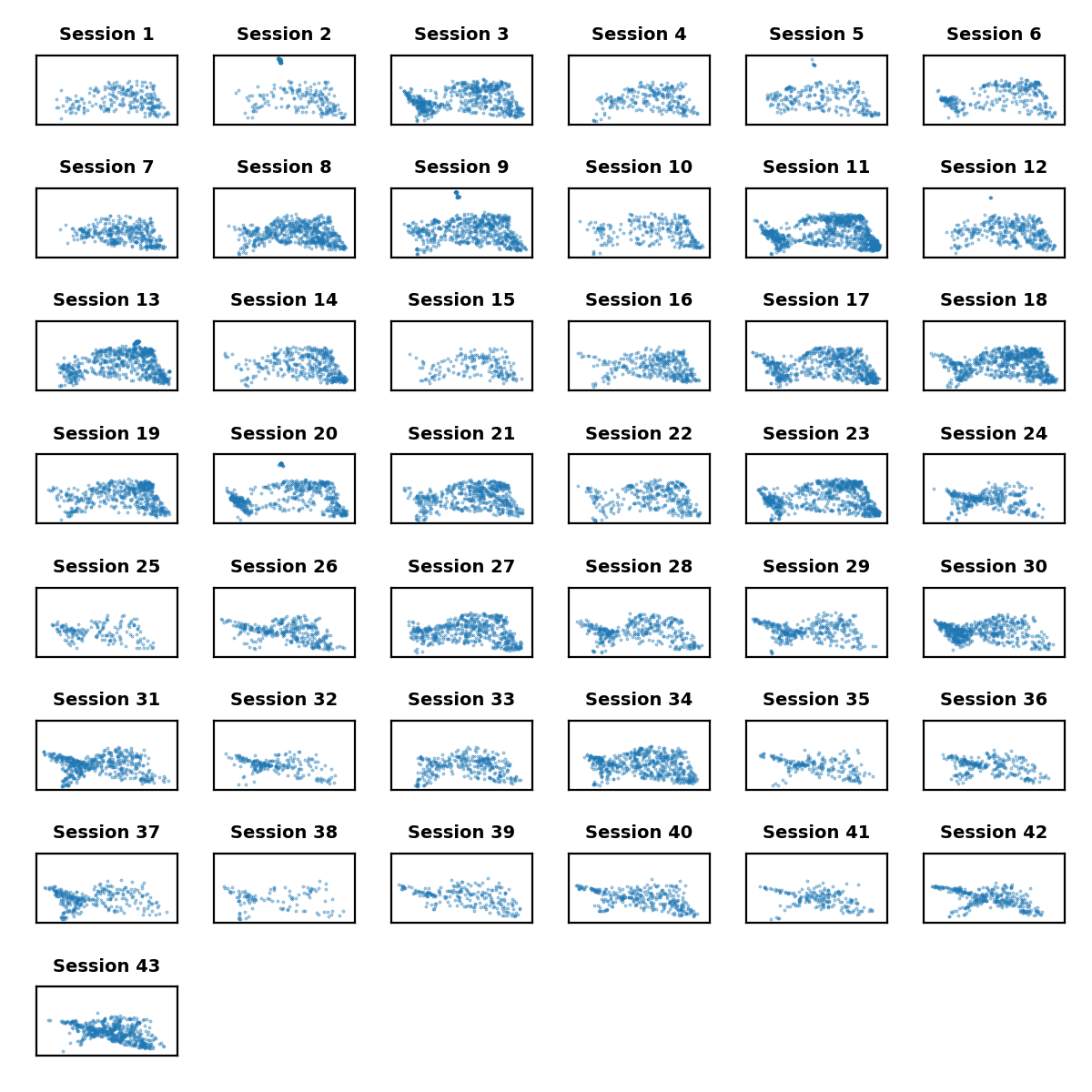}
	\caption{Right hand dataset embedding showed for every session separately.} \label{fig:embedding-rh}
\end{figure*}

\section{Intrinsic dimensionality computed with TwoNN}
\begin{figure*}
	\centering \includegraphics[width=\textwidth]{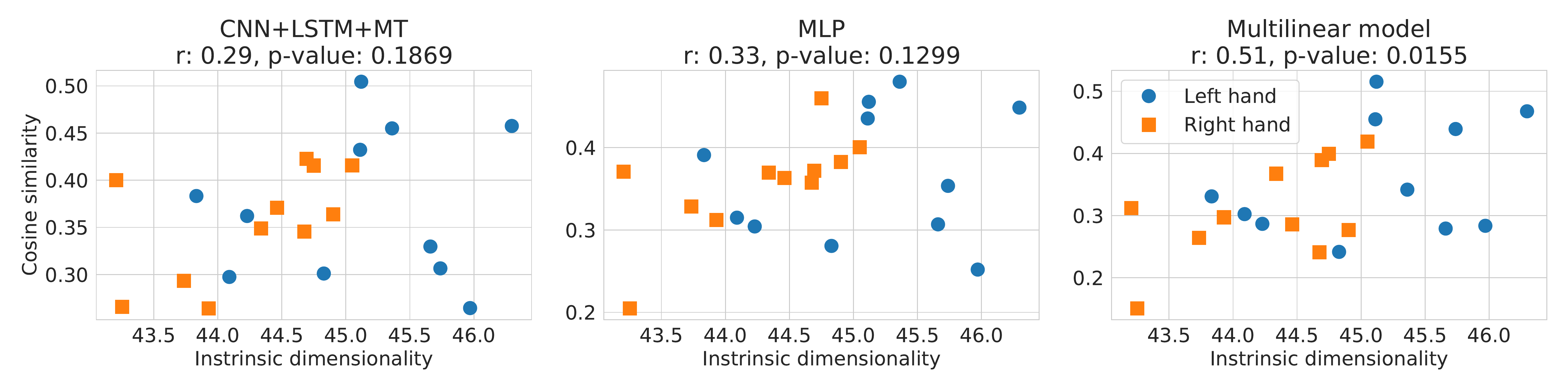}
	\caption{Relationship between cosine similarity and ID of the training dataset computed with TwoNN for dataset translation experiment. In the plot titles, Pearson correlation coefficient r and p-value (the probability of two uncorrelated inputs obtaining r at least as extreme as obtained in this case) are presented.} \label{fig:lid-vs-cs-twonn} 
\end{figure*}

\newpage
\section*{References}
\bibliography{mybibfile}
\end{document}